\documentclass[manuscript,sigconf]{acmart}
\usepackage{multirow}
\usepackage{array}
\usepackage{float}
\usepackage{threeparttable}
\usepackage{algorithm} 
\usepackage{algpseudocode} 
\usepackage{subcaption}  
\usepackage{tabularray}
\usepackage{booktabs}
\usepackage{graphicx}
\usepackage{makecell}

\AtBeginDocument{%
  }

\setcopyright{acmlicensed}
\copyrightyear{2025}
\acmYear{2025}
\acmDOI{XXXXXXX.XXXXXXX}

\acmConference[FSE '25]{Companion Proceedings of the 33rd ACM Symposium on the Foundations of Software Engineering (FSE '25)}{June 23--27, 2025}{Trondheim, Norway}
\acmBooktitle{Companion Proceedings of the 33rd ACM Symposium on the Foundations of Software Engineering (FSE '25), June 23--27, 2025, Trondheim, Norway}

\begin{document}

\title{RUM: Rule+LLM-Based Comprehensive Assessment on Testing Skills}

\author{Yue Wang}
\affiliation{
  \institution{Nanjing University}
  \city{Nanjing}
  \country{China}}

\author{Zhenyu Chen}
\affiliation{
  \institution{Mooctest Inc.}
  \city{Nanjing}
  \country{China}}

\author{Yuan Zhao}
\affiliation{
  \institution{Nanjing University}
  \city{Nanjing}
  \country{China}}

\author{Chunrong Fang}
\affiliation{
  \institution{Nanjing University}
  \city{Nanjing}
  \country{China}}

\author{Ziyuan Wang}
\affiliation{
  \institution{Mooctest Inc.}
  \city{Nanjing}
  \country{China}}

\author{Song Huang}
\affiliation{
  \institution{Mooctest Inc.}
  \city{Nanjing}
  \country{China}}

\begin{abstract}

Over the past eight years, the META method has served as a multidimensional testing skill assessment system in the National College Student Contest on Software Testing, successfully assessing over 100,000 students' testing skills. However, META is primarily limited to the objective assessment of test scripts, lacking the ability to automatically assess subjective aspects such as test case and test report. To address this limitation, this paper proposes RUM, a comprehensive assessment approach that combines rules and large language models (LLMs). RUM achieves a comprehensive assessment by rapidly processing objective indicators through rules while utilizing LLMs for in-depth subjective analysis of test case documents, test scripts, and test reports. The experimental results show that compared to traditional manual testing skill assessment, RUM improves assessment efficiency by 80.77\% and reduces costs by 97.38\%, while maintaining high accuracy and consistency of assessment. By applying RUM on the contest on software testing, we find that it not only enhances the efficiency and scalability of skill assessment in software testing education, but also provides teachers with more comprehensive and objective evidence for student ability assessment, facilitating personalized teaching and learning. This study offers new insights into the assessment of testing skills, which are expected to promote further development in test process optimization and software quality assurance. \\Demo System URL: \url{http://47.98.125.77:8099} \\ Demo Video URL: \url{https://youtu.be/0XO5GcBEQxg}

\end{abstract}



\maketitle

\section{Introduction}
\begin{figure*}[t]
  \centering
  \includegraphics[width=\linewidth]{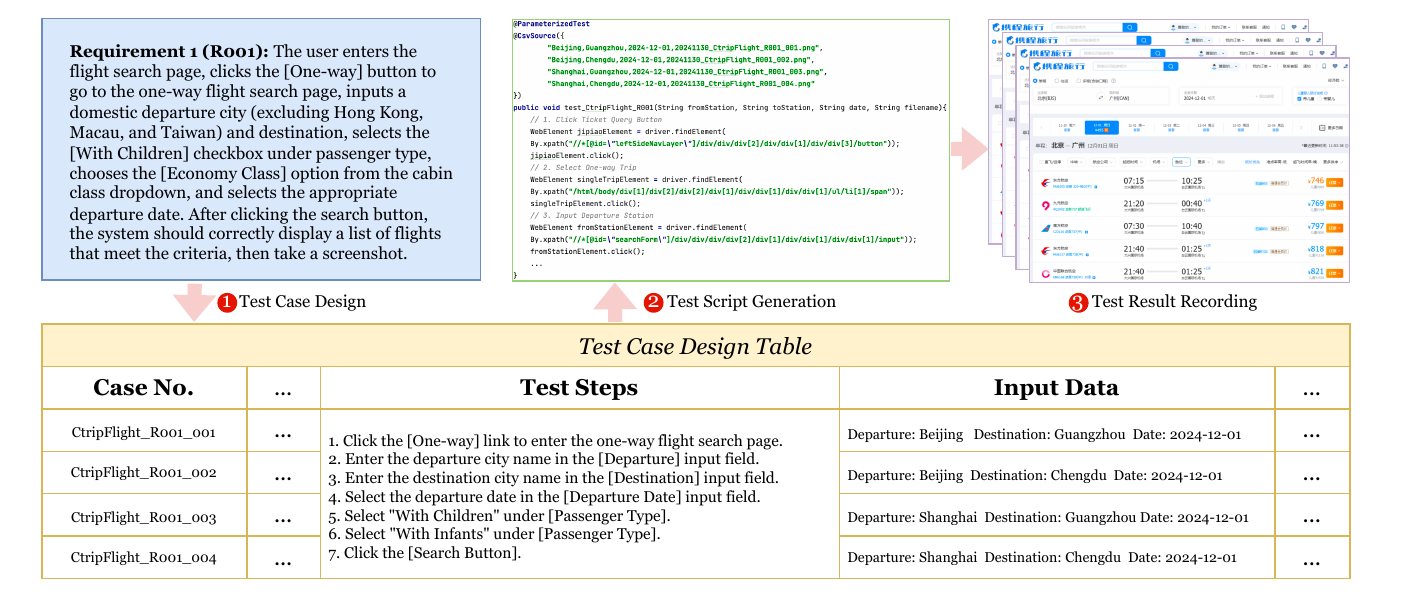}
  \caption{Three Key Steps in the Software Testing Competition Designed to Examine Students' Testing Skills.}
  \label{fig:casestudy}
\end{figure*}
Skill assessment plays a key role in personal development and organizational management. By assessing individual skill levels, organizations can better allocate tasks and improve overall work efficiency \cite{rasul2012employability}. Based on assessment results, the targeted training and development plans can be formulated. Through continuous skill assessment and improvement, both individuals and organizations can maintain a competitive edge. Testing skill assessment can help optimize the testing process and assign appropriate tasks to corresponding testers. Accurate skill assessment can help testers perform testing tasks more quickly and effectively \cite{akdur2022analysis}. The organizations can ensure they have the ability to identify and report software defects, thus guaranteeing software quality\cite{chen2014quasi}.

To assess students' testing skills, we have implemented a multidimensional software testing ability assessment system on MoocTest\footnote{www.mooctest.net}, called META \cite{zhou2022meta}, for the past 8 years of the National College Student Contest on Software Testing in China\cite{wang2019software}. META is used to systematically assess testing skills for large-scale students across seven dimensions for developer testing, web application testing, and mobile application testing. These dimensions are all automated objective assessment methods for test scripts, so that META have assessed over 100,000 students. However, META cannot perform assessment on test requirement analysis, test case design, and test report analysis, because these contents are all subjective in nature making it difficult to implement large-scale automated assessment.

The subjective aspects of testing skill assessment, including test case design quality and test report analysis, require deep understanding and professional judgment. These components focus on assessing whether test cases cover all possible inputs, whether test step descriptions are clear and understandable, whether the code style is standardized, and whether test cases provide sufficient detailed information to ensure test repeatability. Traditional manual assessment of these aspects is time-consuming and often leads to scoring inconsistencies among different assessors due to their flexible assessment criteria and context-dependent feature. Moreover, the increasing scale of testing education and certification programs demands a more efficient and standardized approach to subjective assessment \cite{zhou2022meta}.

While the objective assessment of testing skills focuses on aspects that can be measured through explicit rules and standards, such as test case number naming conventions, module name correctness, syntax error detection, and timestamp format verification. Although these rule-based assessment methods can provide consistent standards and ensure fairness \cite{ullah2019rule}, they may become rigid and fail to adapt to evolving testing practices \cite{reichhart2007rule}. Additionally, when assessment rules become widely known, students might focus on meeting specific metrics rather than developing comprehensive testing skills. These limitations highlight the need for a more flexible and adaptive assessment approach that can maintain objectivity while accommodating the dynamic nature of software testing practices.

To address the challenges in both objective and subjective assessment, we propose \textbf{RUM}, a comprehensive assessment method that combines \textbf{RU}LEs and LL\textbf{M}s, capable of conducting both objective and subjective assessments simultaneously. RUM leverages the consistency of rule-based assessment for objective indicators while utilizing LLMs for in-depth analysis of subjective aspects \cite{askarbekuly2024llm}. This hybrid approach enables automated assessment of both test case design quality and test report analysis, while maintaining the efficient assessment of test scripts established by META.

Our experimental results demonstrate RUM's effectiveness in both accuracy and efficiency. Compared to traditional manual assessment, RUM achieves an 80.77\% improvement in assessment efficiency and a 97.38\% reduction in costs. RUM maintains high assessment accuracy with a correlation coefficient with human assessors, while significantly reducing the standard deviation in scoring consistency. These results indicate that RUM successfully balances the need for objective consistency with subjective comprehension in the assessment of testing skills.

In a comprehensive case study conducted during the final of the 2024 National College Student Contest on Software Testing, RUM was used to assess the testing skills of 148 participants within three hours. The assessment covered test case design, test script implementation, and result analysis for the flight ticket functionality of Ctrip\footnote{www.ctrip.com}. RUM's results were highly correlated with expert assessments (correlation coefficient > 0.8) and received positive feedback from both instructors and students. This real-world application demonstrates RUM's capability to handle large-scale assessment tasks while maintaining assessment quality and reliability.

The noteworthy contributions of this paper can be concluded as follows:

\begin{itemize}
    \item We propose and implement RUM, a novel comprehensive assessment approach that combines rules and LLMs to effectively assess both objective and subjective aspects of software testing skills, addressing the limitations of traditional assessment approaches.
    \item We conduct a experiment from the perspective of consistency, stability, and efficiency, demonstrating the effectiveness of the RUM.
    \item We apply RUM in the National College Student Contest on Software Testing to assess the testing skills of students, demonstrating its capability to enhances the efficiency and scalability of skill assessment in software testing education.

\end{itemize}

\section{Background and Motivation}

\subsection{Skill Requirements of Software Testing}

\label{section2.1}
Software testing includes several stages, each with unique skill requirements. The requirements analysis stage needs accurate interpretation of requirement documents, risk identification, and clear communication with the test team. Test case design requires proficiency in various methods to create comprehensive and maintainable cases. Test script coding demands mastery of automation tools and programming languages to produce stable and efficient scripts. Finally, test report analysis involves recording results clearly and writing comprehensive summaries. These skills are interconnected and form a complete software testing skill set.

The capabilities of these four stages of software testing (requirements analysis, case design, script coding, and report analysis) are interrelated and influence each other. Understanding requirements well directly impacts test case quality. Proficient case design improves script execution and maintainability. Mastery of automation tools affects script stability and efficiency. Accurate reporting helps identify issues and influences future testing cycles. Therefore, skill assessments should consider these interdependencies. Figure \ref{fig:casestudy} illustrates the three key steps abstracted from the aforementioned four stages of the software testing process, which serve as a competition topic to examine the students' testing skills in the 2024 National College Student Contest on Software Testing.

\subsection{Limitation of META}

Although META\cite{zhou2022meta} has become relatively sophisticated in the assessment of test code, its assessment dimensions are confined to objective aspects and are incapable of assessing subjective dimensions. As different students have varying design ideas, and the test cases and reports they create are typically described in natural language and even include some screenshots, the content of these materials mostly requires subjective assessment. As elaborated in Section \ref{section2.1}, the assessment of these two components is a crucial part of the assessment of testing skills. RUM fills the gap in the assessment of these two components, not only being able to conduct subjective assessments of test case documents and test reports but also being capable of assessing the subjective aspects of test code on the basis of META. Harnessing the capabilities of LLMs, RUM can provide a more comprehensive, in-depth, and personalized assessment of testing skills. This not only assesses testers' technical abilities but also assesses their soft skills such as analysis, innovation, and communication, thus better meeting the needs of modern software testing.

\subsection{Rule-based and LLM-based Assessment}
In software testing skill assessment, different methods have emerged to address various assessment needs. Rule-based assessment methods excel at objective assessment, focusing on measurable aspects with clear standards and criteria. These methods efficiently assess aspects such as test case naming conventions, code syntax correctness, and timestamp format verification through predefined rules and automated processes, providing consistency and scalability. However, they often struggle with complex aspects requiring human judgment. Conversely, LLM-based assessment methods demonstrate particular strength in subjective assessment, capable of understanding and assessing aspects that require professional judgment, such as test strategy rationality and problem-solving ability. While LLMs offer flexibility and contextual understanding, they face challenges in consistency and resource efficiency. The complementary nature of these approaches suggests that combining them could address their respective limitations while leveraging their strengths \cite{ullah2019rule, grevisse2024llm}. This combination forms the foundation of our proposed RUM system, which aims to achieve both efficient and comprehensive assessment of software testing skills.

\subsection{Motivation}
Based on our analysis of META's limitations and the characteristics of rule-based and LLM-based assessment methods, there is a clear need for an integrated approach that can comprehensively assess both objective and subjective aspects of software testing skills. While META has successfully assessed over 100,000 students' test scripts through its seven-dimensional framework, it struggles with subjective assessments such as test case design quality and test report analysis. Rule-based methods excel in providing consistent and efficient assessments of objective criteria but lack the flexibility to assess complex aspects like creativity and problem-solving abilities. Conversely, LLM-based methods offer the capability to understand and  assess subjective components but face challenges in consistency and resource efficiency. This motivates us to develop RUM, which combines the strengths of both approaches while mitigating their respective limitations. By integrating rule-based assessment's efficiency and consistency with LLMs' ability to comprehend and assess subjective aspects, RUM aims to provide a more comprehensive, scalable, and reliable assessment solution for software testing education and practice \cite{zhou2022meta, askarbekuly2024llm,chen2023education}.

\section{Approach}
\begin{figure*}[h]
  \centering
  \includegraphics[width=\linewidth]{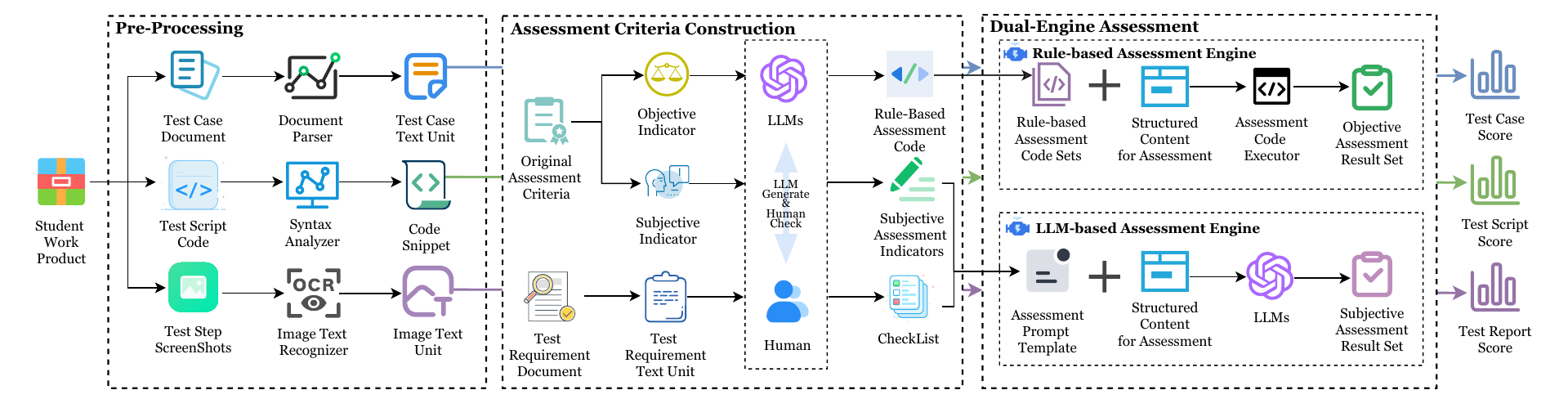}
  \caption{The Framework of RUM.}
  \label{fig:overview}
\end{figure*}

In this section, we present our proposed approach, RUM, in detail. As shown in Figure \ref{fig:overview}, RUM is composed of three stages: pre-processing, assessment criteria construction, and dual-engine assessment. Firstly, the pre-processing stage handles different input formats of test work products, including test case document, test script code, and test step screenshots, and prepares them for assessment. Secondly, the criteria construction stage leverages LLMs to construct rule-based assessment code, subjective assessment indicators, and checklist. Thirdly, the dual-engine assessment stage combines rule-based and LLM-based methods to assess both objective and subjective aspects of the testing skills. This integrated approach enables RUM to generate comprehensive, efficient, and reliable assessment results for large-scale assessment of testing skills.

\subsection{Pre-Processing}
As the first stage of RUM, it is essential to unify the content to be assessed. In the real world, students submit their test products in various forms and these products often contain multiple modalities, so it is crucial to unify them into structured content for assessment to facilitate the subsequent assessment process. In this subsection, we provide a detailed introduction to the preprocessing phase of RUM, which is responsible for handling and standardizing various types of input materials into a unified structured content for assessment.

For \textbf{test case documents}, we implement a \textbf{document parser} that utilizes  docx, openpyxl, and pdfplumber libraries to parse different document formats (Excel, Word, PDF) and convert them into a unified JSON structure, called \textbf{test case text unit}, containing standardized fields such as test case ID, test case description, test steps, expected result, and actual result. For \textbf{test scripts}, we implement a two-phase analysis \textbf{syntax analyzer} combining abstract syntax tree (AST) parsing and regular expression matching. The AST parser analyzes the code structure to identify function definitions, method signatures, and dependencies, while regular expressions extract key information such as parameterized input data and operational steps of the test cases. By integrating these elements, we finally construct our code snippets. For \textbf{screenshots}, we employ Alibaba Cloud's high-concurrency OCR API as the \textbf{image text recognizer} to extract text information from screenshots and use it as the \textbf{image text unit}. The OCR technology can accurately recognize and extract test results from screenshots, thereby avoiding the performance and cost overheads associated with directly using large language models with multimodal capabilities.

The \textbf{test case text unit}, \textbf{code snippet}, and \textbf{image text unit} obtained through the above three preprocessing steps are regarded as the \textbf{structured content for assessment}, which will be utilized in the dual-engine assessment stage by being embedded into prompt templates or fed into a rule-based assessment engine.

\subsection{Assessment Criteria Construction}

\begin{table*}[h]
\caption{Skill Assessment Indicators for Test Work Products}
\label{tab:test-work-products-assessment}
\resizebox{\textwidth}{!}{%
\begin{tabular}{ccclcc}
\toprule
\textbf{Work Product} & \textbf{Indicator} & \textbf{Name} & \textbf{Description} & \textbf{Perspective} & \textbf{Method} \\
\midrule
\multirow{11}{*}{Test Case Document} & \multirow{3}{*}{Standardization} & $STAN_1$ & Is the naming of test case numbers compliant with standards? & Objective & Rule \\
                                     &                                  & $STAN_2$ & Is the module name correct? & Objective & Rule \\
                                     &                                  & $STAN_3$ & Are the descriptions of the test case execution steps standardized? & Objective+Subjective & Rule+LLM \\
\cmidrule{2-6}
                                     & Sufficiency & $SUFF_1$ & Does the test case cover all possible inputs? & Subjective & LLM \\
\cmidrule{2-6}
                                     & \multirow{2}{*}{Completeness} & $COMP_1$ & Is the test case associated with a screenshot? & Objective & Rule \\
                                     &                               & $COMP_2$ & Are all fields of the test case completely filled out? & Objective & Rule \\
\cmidrule{2-6}
                                     & \multirow{2}{*}{Consistency} & $CONS_1$ & Is the description of the operation steps consistent with the logic of the test script? & Subjective & LLM \\
                                     &                              & $CONS_2$ & Are the input data of the test case consistent with the parameterized data in the test script? & Subjective & LLM \\
\cmidrule{2-6}
                                     & \multirow{3}{*}{Readability} & $READ_1$ & Are the test cases clearly describing the purpose, steps, and expected results of the test? & Subjective & LLM \\
                                     &                              & $READ_2$ & Do the test cases follow a consistent structure and format? & Subjective & LLM \\
                                     &                              & $READ_3$ & Do the test cases provide enough detail to ensure the repeatability and accuracy of the tests? & Subjective & LLM \\
\midrule
\multirow{7}{*}{Test Script Code} & \multirow{3}{*}{Standardization} & $STAN_4$ & Does the code include the necessary structure of the test script? & Objective & Rule \\
                                  &                                  & $STAN_5$ & Are there any syntax errors in the code? & Objective & Rule \\
                                  &                                  & $STAN_6$ & Is the coding style standardized? & Subjective & LLM \\
\cmidrule{2-6}
                                  & Consistency(*) & $-$ & This indicator is associated with consistency $CONS_1$ and $CONS_2$ in test case assessment. & $-$ & $-$ \\
\cmidrule{2-6}
                                  & \multirow{2}{*}{Runnability} & $RUNN_1$ & Does the screenshot naming timestamp conform to the required format? & Objective & Rule \\
                                  &                              & $RUNN_2$ & Does the content of the screenshot match the expected results described in the requirements? & Subjective & LLM \\
\cmidrule{2-6}
                                  & Sufficiency & $SUFF_2$ & Does the screenshots cover the results generated by all input data? & Subjective & LLM \\
\midrule
Test Report & $-$ & $-$ & The assessment of the test report is associated with the assessment of test cases and test code. & $-$ & $-$ \\
\bottomrule
\end{tabular}
}
\footnotesize{\textit{* represents common indicators for both Test Case Document and Test Script Code}}
\end{table*}

The second stage of RUM is assessment criteria construction. Any assessment initially requires the establishment of assessment criteria, and the assessment of testing skills is no exception. In the assessment of software testing skills, there are mainly two types of documents related to the assessment criteria. The first one is the \textbf{original assessment criteria}, which are usually the assessment standards set by teachers in the classroom or the competition rules established by the organizing committee in testing competitions. These criteria are often described in natural language and consist of both subjective and objective aspects. The second one is the \textbf{test requirement document}, which is also typically described in natural language and outlines the specific requirements and objectives of the testing tasks, including the expected functionalities, constraints, and key componets to focus on during the testing process.

In the stage of criteria construction, we utilizes these two documents to construct assessment criteria. For \textbf{original assessment criteria}, we first constructed an assessment system based on expert experience and the description of the original assessment criteria, which includes 6 dimensions and 17 assessment indicators, as shown in Table \ref{tab:test-work-products-assessment}. Each indicator is categorized into subjective and objective indicators according to their attributes. For \textbf{objective indicators} (such as naming conventions, field completeness, etc.), we designed specific prompt templates to guide the LLMs to generate \textbf{rule-based assessment code} that can be directly loaded into the rule-based assessment engine. For \textbf{subjective indicators} (such as test case coverage, code readability, etc.), we use LLMs to analyze and refine these subjective indicators described in natural language and transform them into part of the \textbf{subjective assessment indicators} in the assessment prompt template of LLM-based assessment engine. For \textbf{test requirement document}, it is firstly decomposed into \textbf{test requirement text unit} based on pre-set granularity, with each unit containing specific testing requirements description. We then use prompt templates to leverage LLMs to transform the unit into \textbf{checklist}.

After the aforementioned LLM processing and manual review, we obtained the final assessment criteria, including \textbf{rule-based assessment code}, \textbf{subjective assessment indicators}, and \textbf{checklists}. These outputs will serve as the assessment standards in the dual-engine assessment stage, to be used by the rule-based assessment engine and the LLM-based assessment engine.

\subsection{Dual-Engine Assessment}
The dual-engine assessment stage combines rule-based and LLM-based assessment methods to provide comprehensive skill assessment. The rule-based assessment engine is designed with an object-oriented approach. Specifically, it first loads all the rule-based assessment code into the engine, and use the \textbf{assessment code executor} to execute the rule-based assessment code. In detail, the symbolic representation of the rule-based assessment engine is as follows:
\begin{equation}
\mathcal{R}_i(I, S, F) = {E}_{\textit{Rule}}(C, R_i)
\end{equation}

The formula represents the implementation of a rule-based assessment engine, where:

\begin{itemize}
    \item ${E}_{\textit{Rule}}(C, R_i)$ denotes the rule-based assessment engine ${E}_{\textit{Rule}}$, which takes two inputs: \textbf{the structured content for assessment} $C$ and the rule $R_i$(essentially a rule-based assessment code). The rule engine applies the rule set $R_i$ to process the content $C$, thereby generating the final \textbf{objective assessment result set} $\mathcal{R}_i(I, S, F)$.
    \item $\mathcal{R}_i(I, S, F)$ represents the objective assessment result set, which consists of three parts: the indicator $I$, the score $S$, and the feedback $F$. This result set is obtained by processing the content $C$ according to the rule $R_i$.
\end{itemize}

Different rules $R_i$ are applied to process the content $C$, generating objective corresponding assessment results, including the assessment indicators ${I}$, scores ${S}$, and feedback ${F}$. This structured assessment approach enables the rule-based assessment engine to flexibly adapt to the switching of different rules and produce different assessment outputs based on the input rule.

The LLM-based assessment engine is driven by large language models. We first design an example \textbf{assessment prompt template} for the engine, which consists mainly of four parts: \(\textit{Input}\), \(\textit{Rules}\), \(\textit{Criteria}\), and \(\textit{Output}\). The \(\textit{Input}\) embeds the structured content for assessment, incorporating the original test requirement text unit and the student's test products into the prompt. The \(\textit{Rules}\) integrate checklist constraints into the prompt to ensure the stability of the LLM scoring process and results. The \(\textit{Criteria}\) include explanations of the \textbf{subjective assessment indicators} and the deduction rules for the scores. The \(\textit{Output}\) defines the rules for outputting the scoring results, facilitating the engine to extract the results from the LLM's output using regular expressions. These four parts form the main components of our assessment prompt. We have designed three different templates for test case documents, test script code, and test screenshots, to accommodate assessments of different test products. Once all the required content has been filled into the template, the final prompt is obtained. This prompt is then fed into the LLMs, and the \textbf{subjective assessment result set} is extracted from the results returned by the LLMs through the regular expression mechanism. 

Finally, the dual-engine assessment stage integrates and processes both the objective and subjective assessment result sets from the two engines, ultimately outputting the assessment scores for each test product.

\section{Experiment}
To evaluate the effectiveness of RUM, we constructed a dataset collected from MoocTest, which includes answers submitted by 148 students in the finals of the National College Student Contest on Software Testing, and set three research questions based on this dataset.
\subsection{Research Questions}

\begin{itemize}
    \item \textbf{RQ1}: How does RUM perform compared to manual assessment in terms of accuracy and consistency?
    \item \textbf{RQ2}: How does RUM leverage the combination of rules and LLMs to improve both assessment quality and stability compared to rule-based methods?
    \item \textbf{RQ3}: How does RUM balance assessment efficiency and cost-effectiveness compared to traditional manual grading in large-scale test skill assessment scenarios?
\end{itemize}

\subsection{Experiment Results}
\begin{figure}[h]
  \centering
  \includegraphics[width=\linewidth]{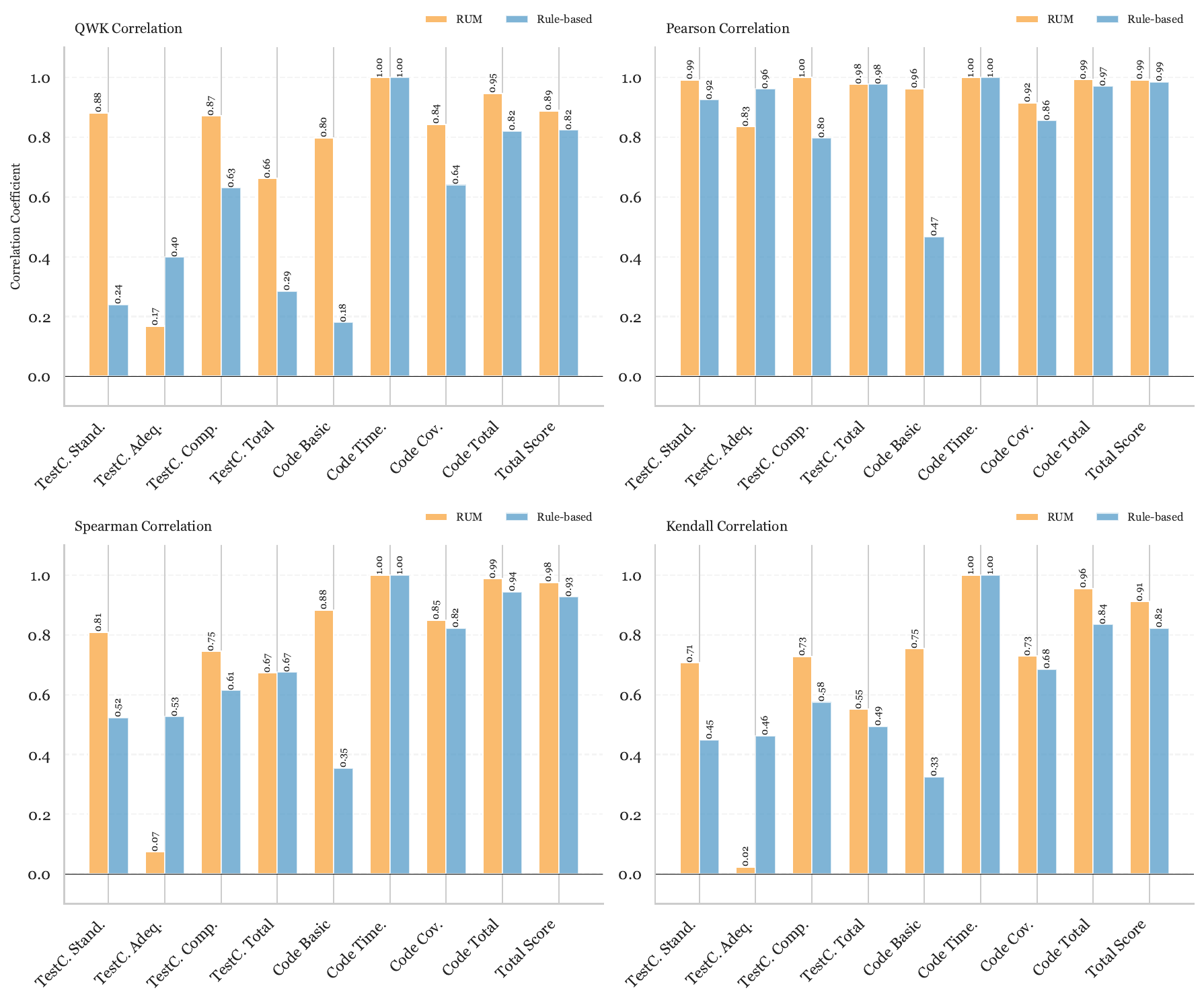}
  \caption{Experiment Results on Correlation Analysis.}
  \label{fig:rq1-fig1}
\end{figure}

\begin{table*}[htbp]
\centering
\caption{Correlation Analysis: Comparing RUM and Rule-based Methods with Human Assessment}
\label{tab:comparison_results}
\resizebox{\textwidth}{!}{
\begin{tabular}{lcccccccccc}
\toprule
\multirow{2}{*}{\textbf{Metric}} & \multicolumn{5}{c}{\textbf{RUM vs Human}} & \multicolumn{5}{c}{\textbf{Rule-based vs Human}} \\
\cmidrule(lr){2-6} \cmidrule(lr){7-11}
& \textbf{Kendall} & \textbf{MAE} & \textbf{Pearson} & \textbf{QWK} & \textbf{Spearman} & \textbf{Kendall} & \textbf{MAE} & \textbf{Pearson} & \textbf{QWK} & \textbf{Spearman} \\
\midrule
\textbf{Basics} & \textbf{0.754} & \textbf{1.40} & \textbf{0.962} & \textbf{0.798} & \textbf{0.883} & 0.325 & 5.70 & 0.467 & 0.180 & 0.355 \\
\textbf{Coverage} & \textbf{0.730} & \textbf{1.40} & \textbf{0.915} & \textbf{0.844} & \textbf{0.849} & 0.685 & 3.70 & 0.856 & 0.641 & 0.823 \\
\textbf{Timestamp} & \textbf{1.000} & \textbf{0.00} & \textbf{1.000} & \textbf{1.000} & \textbf{1.000} & \textbf{1.000} & \textbf{0.00} & \textbf{1.000} & \textbf{1.000} & \textbf{1.000} \\
\textbf{Code Total} & \textbf{0.956} & \textbf{2.10} & \textbf{0.993} & \textbf{0.947} & \textbf{0.988} & 0.836 & 9.40 & 0.971 & 0.821 & 0.944 \\
\midrule
\textbf{Standardization} & \textbf{0.706} & \textbf{1.10} & \textbf{0.992} & \textbf{0.881} & \textbf{0.809} & 0.450 & 10.88 & 0.925 & 0.240 & 0.523 \\
\textbf{Adequacy} & 0.024 & 3.45 & 0.835 & 0.168 & 0.074 & \textbf{0.463} & \textbf{1.85} & \textbf{0.962} & \textbf{0.399} & \textbf{0.527} \\
\textbf{Completeness} & \textbf{0.728} & \textbf{0.05} & \textbf{0.999} & \textbf{0.872} & \textbf{0.745} & 0.576 & 2.75 & 0.797 & 0.630 & 0.614 \\
\textbf{Test Case Total} & \textbf{0.552} & \textbf{4.30} & \textbf{0.978} & \textbf{0.662} & 0.673 & 0.494 & 11.78 & 0.977 & 0.285 & \textbf{0.675} \\
\midrule
\textbf{Total Score} & \textbf{0.911} & \textbf{5.50} & \textbf{0.992} & \textbf{0.889} & \textbf{0.976} & 0.822 & 7.04 & 0.985 & 0.824 & 0.927 \\
\bottomrule
\end{tabular}
}
\end{table*}
\begin{table*}[h]
\centering
\caption{Score Range Analysis of RUM Assessments Across Five Students}
\label{tab:stability_analysis}
\resizebox{\textwidth}{!}{
\begin{tabular}{cccccccccc}
\toprule
\textbf{Student} & \textbf{Standardization} & \textbf{Adequacy} & \textbf{Completeness} & \textbf{Basics} & \textbf{Timestamp} & \textbf{Coverage} & \textbf{TestCase Total} & \textbf{Code Total} & \textbf{Total Score} \\
\midrule
Stu1 & 47.0-50.0 & 20.0-24.0 & 26.0 & 23.5-24.5 & 39.0 & 25.0 & 94.0-100.0 & 87.5-88.5 & 182.5-188.5 \\
Stu2 & 49.0 & 23.0-24.0 & 26.0 & 21.5-22.0 & 39.0 & 20.0 & 98.0-99.0 & 80.5-81.0 & 178.5-180.0 \\
Stu3 & 44.0-46.0 & 22.0-24.0 & 26.0 & 22.0-24.0 & 39.0 & 23.0 & 92.0-96.0 & 84.0-85.5 & 176.0-180.5 \\
Stu4 & 44.0-47.0 & 23.0-24.0 & 26.0 & 18.5-20.5 & 39.0 & 19.0 & 93.0-96.0 & 76.5-78.5 & 170.5-174.5 \\
Stu5 & 35.0-39.0 & 18.0-21.0 & 26.0 & 22.5-23.0 & 39.0 & 25.0 & 79.0-86.0 & 86.5-87.0 & 166.0-173.0 \\
\midrule
\textbf{Max Range} & 4.0 & 3.0 & - & 2.0 & - & - & 7.0 & 2.0 & 7.0 \\
\textbf{Avg Range} & 2.4 & 1.8 & - & 1.5 & - & - & 4.2 & 1.4 & 4.8 \\
\bottomrule
\end{tabular}
}
\end{table*}

\textbf{Answer to RQ1}: Our experimental results demonstrate that RUM achieves high agreement with human assessment across multiple evaluation metrics. As shown in Table \ref{tab:comparison_results}, RUM maintains strong consistency with human assessors, achieving a Quadratic Weighted Kappa (QWK) of 0.889 for total scores, indicating excellent inter-rater reliability. The approach performs particularly well in assessing code-related aspects, with correlation coefficients above 0.9 for code total scores. For test case assessment, RUM shows strong performance in standardization (QWK: 0.881) and completeness (QWK: 0.872). The mean absolute error (MAE) remains consistently low across different assessment aspects, with most metrics showing MAE values below 2.0. This indicates that RUM can effectively replicate human judgment while maintaining consistency across different assessment dimensions.

\textbf{Answer to RQ2}: The organic combination of rules and LLMs in RUM brings two significant improvements: enhanced assessment quality compared to rule-based methods and improved LLM-based assessment stability through rule constraints. First, regarding assessment quality, as shown in Figure \ref{fig:rq1-fig1} and Table \ref{tab:comparison_results}, RUM demonstrates substantial improvements across multiple metrics compared to rule-based methods. For code basics evaluation, RUM reduces the mean absolute error by 75.4\% (MAE: 1.40 vs. 5.70) while achieving stronger correlations with human judgment (Spearman: 0.883 vs. 0.355). In test case standardization assessment, RUM shows particular strength (QWK: 0.881 vs. 0.240), demonstrating the value of LLM's natural language understanding capabilities. Second, regarding stability enhancement through rule combination, our analysis of LLM uncertainty involved 25 experiments across five students' work products, with results shown in Table \ref{tab:stability_analysis}. For purely rule-governed metrics like completeness and timestamp verification, RUM achieves perfect stability (range: 0.0), while metrics requiring LLM judgment maintain controlled variation through external checklist constraints - code basics shows low fluctuation (maximum range: 2.0, average range: 1.5) despite diverse implementation approaches, and standardization shows slightly higher but acceptable variation (maximum range: 4.0, average range: 2.4). The total score fluctuation (average range: 4.8) and overall performance metrics (QWK: 0.889 vs. 0.824, MAE: 5.50 vs. 7.04) demonstrate that RUM successfully balances LLM flexibility with rule-based stability.

\begin{table}[h]
\centering
\caption{Comprehensive Efficiency Analysis of RUM vs. Manual Assessment}
\label{tab:efficiency_analysis}
\footnotesize
\setlength{\tabcolsep}{3.5pt}
\resizebox{0.475\textwidth}{!}{%
\begin{tabular}{@{}lrrrrrrrr@{}}
\toprule
\multirow{2}{*}{\textbf{Method}} &
\multicolumn{2}{c}{\textbf{Time (mm:ss)}} & 
\multicolumn{2}{c}{\textbf{Tokens}} & 
\multicolumn{3}{c}{\textbf{Cost (\$/sub)}} & 
\textbf{Capacity\textsuperscript{*}} \\ 
\cmidrule(lr){2-3} \cmidrule(lr){4-5} \cmidrule(lr){6-8} \cmidrule(lr){9-9}
& Avg. & Std.Dev. & In & Out & Labor & API & Total & (sub/day) \\
\midrule
Manual & 21:17.61 & 01:49.24 & -- & -- & 7.10 & -- & 7.10 & 22.56 \\
RUM & 04:05.71 & 00:29.83 & 713.88 & 240.53 & 0.17 & 0.016 & 0.186 & 351.84 \\
\midrule
\multicolumn{1}{r}{Improvement} & 
\multicolumn{1}{r}{80.77\%} & 
\multicolumn{1}{r}{72.69\%} & 
\multicolumn{2}{r}{954.41 total} & 
\multicolumn{1}{r}{97.61\%} & 
\multicolumn{1}{r}{--} & 
\multicolumn{1}{r}{97.38\%} & 
\multicolumn{1}{r}{1459.13\%} \\
\bottomrule
\multicolumn{9}{@{}l}{\scriptsize \textsuperscript{*}Manual: 8-hour/day, RUM: 24-hour parallel} \\
\multicolumn{9}{@{}l}{\scriptsize Model: moonshot-v1-8k\footnote{www.moonshot.cn}, API Rate: \$1.67/1M tokens (¥12/1M tokens), Labor Rate: \$20/hour)}
\end{tabular}}
\end{table}

\textbf{Answer to RQ3}: The experimental results demonstrate that RUM significantly outperforms traditional manual assessment in both efficiency and cost-effectiveness. As shown in Table \ref{tab:efficiency_analysis}, RUM reduces the average processing time by 80.77\% (from 21:17 to 4:05) while maintaining higher consistency (standard deviation reduced by 72.69\%). The system processes each submission with an average of 954.41 tokens, resulting in minimal API costs (\$0.016 per submission). Combined with reduced human supervision costs, RUM achieves a 97.38\% reduction in total cost per submission (from \$7.10 to \$0.186) while increasing daily processing capacity by over 14 times (from 22.56 to 351.84 submissions), making it highly suitable for large-scale assessment scenarios.

\section{Related Work}

\subsection{Objective and Subjective Assessment}
Objective and subjective assessments offer unique advantages. Mohammadi et al. reviewed image quality methods, emphasizing combined evaluations\cite{mohammadi2014subjective}. Machine learning tools gain traction; Lam et al. noted their accuracy in surgical assessments\cite{lam2022machine}. Tian et al. used recruitment data and NLP for software requirements assessment\cite{tian2022software}. Hettiarachchi et al. developed an engineering education assessment system\cite{hettiarachchi2015assessment}. Cammaerts et al. explored MBT's impact using TesCaV\cite{cammaerts2024assessing}. Ricchiardi et al. assessed soft skills via projects\cite{ricchiardi2018soft}. These studies highlight e-learning's potential in skill assessment, with room for optimization.

\subsection{LLM-based Assessment \& LLM-as-a-Judge}

With the advancement of natural language processing, pre-trained models like BERT\cite{devlin2018bert} and GPT\cite{radford2018improving} have shown excellence in automated text evaluation. Sun et al. proposed MAF for detecting plagiarism in test code, potentially using large language models\cite{sun2019maf}. A systematic review surveyed large language models' applications in computer science education\cite{raihan2024llm}. Other studies explored the impact of software testing competitions on education and practice\cite{shi2019el}, and data-driven methods for assessing software engineering skills\cite{buckley2017data}. ASAG\cite{song2024automated} and AES\cite{kostic2024llms} are key areas of research. Feng et al. examined large language models in Chinese AES\cite{feng2024leveraging}. Tang et al. found prompt engineering enhances scoring reliability in AES\cite{tang2024harnessing}. Chang et al. evaluated ChatGPT in ASAG, noting GPT-4's superiority but calling for further research\cite{chang2024automatic}. Wieser et al. investigated ChatGPT's potential in supporting programming education and demonstrated its value for teacher assessment and student guidance \cite{wieser2023investigating}. Daun et al. discussed how ChatGPT could enhance software engineering education through personalized feedback and curriculum adaptation \cite{daun2023chatgpt}. Kohlbacher et al. analyzed common code quality issues among novice programmers to improve teaching strategies \cite{kohlbacher2023common}. Hellas et al. explored the evolving landscape of generative AI in computing education \cite{macneil2024discussing} and discussed leveraging LLMs for next-generation educational technologies \cite{heffernan2024leveraging}. Balse et al. evaluated LLM-generated explanations for logical errors in student programs, finding them comparable to peer-generated explanations \cite{balse2023evaluating}. Tenbergen et al. developed a tool to facilitate calibrated peer reviews in software engineering education \cite{tenbergen2024tool}.

\subsection{Software Testing Education}
In recent years, researchers have explored effective software testing teaching methods. Cheiran et al. used Problem-Based Learning (PBL) to enhance students' practical problem-solving skills\cite{cheiran2017problem}. Deng et al. employed open-source software to reduce costs and provide practical cases\cite{deng2020teaching}. Shi et al. proposed an e-learning and CDIO model-based training method to boost technical skills and learning interest\cite{shi2019research}. Towey et al. applied Metamorphic Testing to stimulate creativity and improve effectiveness\cite{towey2015teaching}. Arcuri Andrea et al. integrated software testing into algorithm courses using GitHub for materials\cite{arcuri2020teaching}. Fraser Gordon et al. explored using the Code Defenders game in education\cite{fraser2020teaching}. CHEN et al. introduced crowdsourced testing in classrooms, enhancing practical effects with enterprise projects\cite{chen2014quasi}. These studies focus on improving teaching methods and student learning outcomes in software testing, providing valuable resources and insights.

\section{Conclusion}

The RUM approach proposed in this study offers an innovative solution to the field of software testing skill assessment. By combining rule-based methods for objective assessment with LLM-based methods for subjective assessment, RUM effectively addresses many challenges in assessing testing skills for large-scale personnel. It also demonstrates a high correlation with human assessment, enhancing both efficiency and cost-effectiveness, and shows remarkable potential to transform software testing education and certification.

\balance
\bibliographystyle{ACM-Reference-Format}
\bibliography{ref.bib}


\begin{thebibliography}{40}

%%% the Web address.

\ifx \showCODEN    \undefined \def \showCODEN     #1{\unskip}     \fi
\ifx \showDOI      \undefined \def \showDOI       #1{#1}\fi
\ifx \showISBNx    \undefined \def \showISBNx     #1{\unskip}     \fi
\ifx \showISBNxiii \undefined \def \showISBNxiii  #1{\unskip}     \fi
\ifx \showISSN     \undefined \def \showISSN      #1{\unskip}     \fi
\ifx \showLCCN     \undefined \def \showLCCN      #1{\unskip}     \fi
\ifx \shownote     \undefined \def \shownote      #1{#1}          \fi
\ifx \showarticletitle \undefined \def \showarticletitle #1{#1}   \fi
\ifx \showURL      \undefined \def \showURL       {\relax}        \fi
% invisible to TeX
\providecommand\bibfield[2]{#2}
\providecommand\bibinfo[2]{#2}
\providecommand\natexlab[1]{#1}
\providecommand\showeprint[2][]{arXiv:#2}

\bibitem[Akdur(2022)]%
        {akdur2022analysis}
\bibfield{author}{\bibinfo{person}{Deniz Akdur}.}
  \bibinfo{year}{2022}\natexlab{}.
\newblock \showarticletitle{Analysis of software engineering skills gap in the
  industry}.
\newblock \bibinfo{journal}{\emph{ACM Transactions on Computing Education}}
  \bibinfo{volume}{23}, \bibinfo{number}{1} (\bibinfo{year}{2022}),
  \bibinfo{pages}{1--28}.
\newblock


\bibitem[Arcuri(2020)]%
        {arcuri2020teaching}
\bibfield{author}{\bibinfo{person}{Andrea Arcuri}.}
  \bibinfo{year}{2020}\natexlab{}.
\newblock \showarticletitle{Teaching Software Testing in an Algorithms and Data
  Structures Course}. In \bibinfo{booktitle}{\emph{2020 IEEE International
  Conference on Software Testing, Verification and Validation Workshops
  (ICSTW)}}. IEEE, \bibinfo{pages}{419--424}.
\newblock


\bibitem[Askarbekuly and Ani{\v{c}}i{\'c}(2024)]%
        {askarbekuly2024llm}
\bibfield{author}{\bibinfo{person}{Nursultan Askarbekuly} {and}
  \bibinfo{person}{Nenad Ani{\v{c}}i{\'c}}.} \bibinfo{year}{2024}\natexlab{}.
\newblock \showarticletitle{LLM examiner: automating assessment in informal
  self-directed e-learning using ChatGPT}.
\newblock \bibinfo{journal}{\emph{Knowledge and Information Systems}}
  (\bibinfo{year}{2024}), \bibinfo{pages}{1--18}.
\newblock


\bibitem[Balse et~al\mbox{.}(2023)]%
        {balse2023evaluating}
\bibfield{author}{\bibinfo{person}{Rishabh Balse}, \bibinfo{person}{Viraj
  Kumar}, \bibinfo{person}{Prajish Prasad}, {and}
  \bibinfo{person}{Jayakrishnan~Madathil Warriem}.}
  \bibinfo{year}{2023}\natexlab{}.
\newblock \showarticletitle{Evaluating the Quality of LLM-Generated
  Explanations for Logical Errors in CS1 Student Programs}. In
  \bibinfo{booktitle}{\emph{Proceedings of the 16th Annual ACM India Compute
  Conference}}. \bibinfo{pages}{49--54}.
\newblock


\bibitem[Buckley and Buckley(2017)]%
        {buckley2017data}
\bibfield{author}{\bibinfo{person}{Ingrid~A Buckley} {and}
  \bibinfo{person}{Winston~S Buckley}.} \bibinfo{year}{2017}\natexlab{}.
\newblock \showarticletitle{Teaching software testing using data structures}.
\newblock \bibinfo{journal}{\emph{International Journal of Advanced Computer
  Science and Applications}} \bibinfo{volume}{8}, \bibinfo{number}{4}
  (\bibinfo{year}{2017}).
\newblock


\bibitem[Cammaerts and Snoeck(2024)]%
        {cammaerts2024assessing}
\bibfield{author}{\bibinfo{person}{Felix Cammaerts} {and}
  \bibinfo{person}{Monique Snoeck}.} \bibinfo{year}{2024}\natexlab{}.
\newblock \showarticletitle{Assessing the testing skills transfer of
  model-based testing on testing skill acquisition}.
\newblock \bibinfo{journal}{\emph{Software and Systems Modeling}}
  (\bibinfo{year}{2024}), \bibinfo{pages}{1--19}.
\newblock


\bibitem[Chang and Ginter(2024)]%
        {chang2024automatic}
\bibfield{author}{\bibinfo{person}{Li-Hsin Chang} {and} \bibinfo{person}{Filip
  Ginter}.} \bibinfo{year}{2024}\natexlab{}.
\newblock \showarticletitle{Automatic Short Answer Grading for Finnish with
  ChatGPT}. In \bibinfo{booktitle}{\emph{Proceedings of the AAAI Conference on
  Artificial Intelligence}}, Vol.~\bibinfo{volume}{38}.
  \bibinfo{pages}{23173--23181}.
\newblock


\bibitem[Cheiran et~al\mbox{.}(2017)]%
        {cheiran2017problem}
\bibfield{author}{\bibinfo{person}{Jean Felipe~P Cheiran},
  \bibinfo{person}{Elder de M.~Rodrigues}, \bibinfo{person}{Ewerson~Luiz de
  S.~Carvalho}, {and} \bibinfo{person}{Jo{\~a}o Pablo~S da Silva}.}
  \bibinfo{year}{2017}\natexlab{}.
\newblock \showarticletitle{Problem-based learning to align theory and practice
  in software testing teaching}. In \bibinfo{booktitle}{\emph{Proceedings of
  the XXXI Brazilian Symposium on Software Engineering}}.
  \bibinfo{pages}{328--337}.
\newblock


\bibitem[Chen(2023)]%
        {chen2023education}
\bibfield{author}{\bibinfo{person}{Zhenyu Chen}.}
  \bibinfo{year}{2023}\natexlab{}.
\newblock \showarticletitle{Education Reform of Software Engineering in the Age
  of AI: Keynote Address}. In \bibinfo{booktitle}{\emph{2023 IEEE/ACIS 23rd
  International Conference on Computer and Information Science (ICIS)}}. IEEE,
  \bibinfo{pages}{2--2}.
\newblock


\bibitem[Chen and Luo(2014)]%
        {chen2014quasi}
\bibfield{author}{\bibinfo{person}{Zhenyu Chen} {and} \bibinfo{person}{Bin
  Luo}.} \bibinfo{year}{2014}\natexlab{}.
\newblock \showarticletitle{Quasi-crowdsourcing testing for educational
  projects}. In \bibinfo{booktitle}{\emph{Companion Proceedings of the 36th
  International Conference on Software Engineering}}.
  \bibinfo{pages}{272--275}.
\newblock


\bibitem[Daun and Brings(2023)]%
        {daun2023chatgpt}
\bibfield{author}{\bibinfo{person}{Marian Daun} {and} \bibinfo{person}{Jennifer
  Brings}.} \bibinfo{year}{2023}\natexlab{}.
\newblock \showarticletitle{How ChatGPT will change software engineering
  education}. In \bibinfo{booktitle}{\emph{Proceedings of the 2023 Conference
  on Innovation and Technology in Computer Science Education V. 1}}.
  \bibinfo{pages}{110--116}.
\newblock


\bibitem[Deng et~al\mbox{.}(2020)]%
        {deng2020teaching}
\bibfield{author}{\bibinfo{person}{Lin Deng}, \bibinfo{person}{Josh Dehlinger},
  {and} \bibinfo{person}{Suranjan Chakraborty}.}
  \bibinfo{year}{2020}\natexlab{}.
\newblock \showarticletitle{Teaching software testing with free and open source
  software}. In \bibinfo{booktitle}{\emph{2020 IEEE International Conference on
  Software Testing, Verification and Validation Workshops (ICSTW)}}. IEEE,
  \bibinfo{pages}{412--418}.
\newblock


\bibitem[Devlin et~al\mbox{.}(2018)]%
        {devlin2018bert}
\bibfield{author}{\bibinfo{person}{Jacob Devlin}, \bibinfo{person}{Ming-Wei
  Chang}, \bibinfo{person}{Kenton Lee}, {and} \bibinfo{person}{Kristina
  Toutanova}.} \bibinfo{year}{2018}\natexlab{}.
\newblock \showarticletitle{BERT: Pre-training of Deep Bidirectional
  Transformers for Language Understanding}.
\newblock \bibinfo{journal}{\emph{arXiv preprint arXiv:1810.04805}}
  (\bibinfo{year}{2018}).
\newblock


\bibitem[Feng et~al\mbox{.}(2024)]%
        {feng2024leveraging}
\bibfield{author}{\bibinfo{person}{Haiyue Feng}, \bibinfo{person}{Sixuan Du},
  \bibinfo{person}{Gaoxia Zhu}, \bibinfo{person}{Yan Zou},
  \bibinfo{person}{Poh~Boon Phua}, \bibinfo{person}{Yuhong Feng},
  \bibinfo{person}{Haoming Zhong}, \bibinfo{person}{Zhiqi Shen}, {and}
  \bibinfo{person}{Siyuan Liu}.} \bibinfo{year}{2024}\natexlab{}.
\newblock \showarticletitle{Leveraging Large Language Models for Automated
  Chinese Essay Scoring}. In \bibinfo{booktitle}{\emph{International Conference
  on Artificial Intelligence in Education}}. Springer,
  \bibinfo{pages}{454--467}.
\newblock


\bibitem[Fraser et~al\mbox{.}(2020)]%
        {fraser2020teaching}
\bibfield{author}{\bibinfo{person}{Gordon Fraser}, \bibinfo{person}{Alessio
  Gambi}, {and} \bibinfo{person}{Jos{\'e}~Miguel Rojas}.}
  \bibinfo{year}{2020}\natexlab{}.
\newblock \showarticletitle{Teaching software testing with the code defenders
  testing game: Experiences and improvements}. In
  \bibinfo{booktitle}{\emph{2020 IEEE International Conference on Software
  Testing, Verification and Validation Workshops (ICSTW)}}. IEEE,
  \bibinfo{pages}{461--464}.
\newblock


\bibitem[Gr{\'e}visse(2024)]%
        {grevisse2024llm}
\bibfield{author}{\bibinfo{person}{Christian Gr{\'e}visse}.}
  \bibinfo{year}{2024}\natexlab{}.
\newblock \showarticletitle{LLM-based automatic short answer grading in
  undergraduate medical education}.
\newblock \bibinfo{journal}{\emph{BMC Medical Education}} \bibinfo{volume}{24},
  \bibinfo{number}{1} (\bibinfo{year}{2024}), \bibinfo{pages}{1060}.
\newblock


\bibitem[Heffernan et~al\mbox{.}(2024)]%
        {heffernan2024leveraging}
\bibfield{author}{\bibinfo{person}{Neil Heffernan}, \bibinfo{person}{Rose
  Wang}, \bibinfo{person}{Christopher MacLellan}, \bibinfo{person}{Arto
  Hellas}, \bibinfo{person}{Chenglu Li}, \bibinfo{person}{Candace Walkington},
  \bibinfo{person}{Joshua Littenberg-Tobias}, \bibinfo{person}{David Joyner},
  \bibinfo{person}{Steven Moore}, \bibinfo{person}{Adish Singla},
  {et~al\mbox{.}}} \bibinfo{year}{2024}\natexlab{}.
\newblock \showarticletitle{Leveraging Large Language Models for
  Next-Generation Educational Technologies}. In
  \bibinfo{booktitle}{\emph{Proceedings of the 17th International Conference on
  Educational Data Mining}}. \bibinfo{pages}{1037--1039}.
\newblock


\bibitem[Hettiarachchi et~al\mbox{.}(2015)]%
        {hettiarachchi2015assessment}
\bibfield{author}{\bibinfo{person}{Enosha Hettiarachchi},
  \bibinfo{person}{M~Antonia Huertas}, {and} \bibinfo{person}{Enric Mor}.}
  \bibinfo{year}{2015}\natexlab{}.
\newblock \showarticletitle{E-assessment system for skill and knowledge
  assessment in computer engineering education}.
\newblock \bibinfo{journal}{\emph{International Journal of Engineering
  Education}} \bibinfo{volume}{31}, \bibinfo{number}{2} (\bibinfo{year}{2015}),
  \bibinfo{pages}{529--540}.
\newblock


\bibitem[Kohlbacher et~al\mbox{.}(2023)]%
        {kohlbacher2023common}
\bibfield{author}{\bibinfo{person}{Christina~Julia Kohlbacher},
  \bibinfo{person}{Michael Vierhauser}, {and} \bibinfo{person}{Iris Groher}.}
  \bibinfo{year}{2023}\natexlab{}.
\newblock \showarticletitle{Common code quality issues of novice Java
  programmers: a comprehensive analysis of student assignments}. In
  \bibinfo{booktitle}{\emph{15th International Conference on Computer Supported
  Education}}.
\newblock


\bibitem[Kostic et~al\mbox{.}(2024)]%
        {kostic2024llms}
\bibfield{author}{\bibinfo{person}{Milan Kostic},
  \bibinfo{person}{Hans~Friedrich Witschel}, \bibinfo{person}{Knut Hinkelmann},
  {and} \bibinfo{person}{Maja Spahic-Bogdanovic}.}
  \bibinfo{year}{2024}\natexlab{}.
\newblock \showarticletitle{LLMs in Automated Essay Evaluation: A Case Study}.
  In \bibinfo{booktitle}{\emph{Proceedings of the AAAI Symposium Series}},
  Vol.~\bibinfo{volume}{3}. \bibinfo{pages}{143--147}.
\newblock


\bibitem[Lam et~al\mbox{.}(2022)]%
        {lam2022machine}
\bibfield{author}{\bibinfo{person}{Kyle Lam}, \bibinfo{person}{Junhong Chen},
  \bibinfo{person}{Zeyu Wang}, \bibinfo{person}{Fahad~M Iqbal},
  \bibinfo{person}{Ara Darzi}, \bibinfo{person}{Benny Lo},
  \bibinfo{person}{Sanjay Purkayastha}, {and} \bibinfo{person}{James~M
  Kinross}.} \bibinfo{year}{2022}\natexlab{}.
\newblock \showarticletitle{Machine learning for technical skill assessment in
  surgery: a systematic review}.
\newblock \bibinfo{journal}{\emph{NPJ digital medicine}} \bibinfo{volume}{5},
  \bibinfo{number}{1} (\bibinfo{year}{2022}), \bibinfo{pages}{24}.
\newblock


\bibitem[MacNeil et~al\mbox{.}(2024)]%
        {macneil2024discussing}
\bibfield{author}{\bibinfo{person}{Stephen MacNeil}, \bibinfo{person}{Juho
  Leinonen}, \bibinfo{person}{Paul Denny}, \bibinfo{person}{Natalie Kiesler},
  \bibinfo{person}{Arto Hellas}, \bibinfo{person}{James Prather},
  \bibinfo{person}{Brett~A Becker}, \bibinfo{person}{Michel Wermelinger}, {and}
  \bibinfo{person}{Karen Reid}.} \bibinfo{year}{2024}\natexlab{}.
\newblock \showarticletitle{Discussing the Changing Landscape of Generative AI
  in Computing Education}. In \bibinfo{booktitle}{\emph{Proceedings of the 55th
  ACM Technical Symposium on Computer Science Education V. 2}}.
  \bibinfo{pages}{1916--1916}.
\newblock


\bibitem[Mohammadi et~al\mbox{.}(2014)]%
        {mohammadi2014subjective}
\bibfield{author}{\bibinfo{person}{Pedram Mohammadi}, \bibinfo{person}{Abbas
  Ebrahimi-Moghadam}, {and} \bibinfo{person}{Shahram Shirani}.}
  \bibinfo{year}{2014}\natexlab{}.
\newblock \showarticletitle{Subjective and objective quality assessment of
  image: A survey}.
\newblock \bibinfo{journal}{\emph{arXiv preprint arXiv:1406.7799}}
  (\bibinfo{year}{2014}).
\newblock


\bibitem[Radford et~al\mbox{.}(2018)]%
        {radford2018improving}
\bibfield{author}{\bibinfo{person}{Alec Radford}, \bibinfo{person}{Karthik
  Narasimhan}, \bibinfo{person}{Tim Salimans}, {and} \bibinfo{person}{Ilya
  Sutskever}.} \bibinfo{year}{2018}\natexlab{}.
\newblock \showarticletitle{Improving Language Understanding by Generative
  Pre-Training}.
\newblock \bibinfo{journal}{\emph{arXiv preprint arXiv:1801.06146}}
  (\bibinfo{year}{2018}).
\newblock


\bibitem[Raihan et~al\mbox{.}(2024)]%
        {raihan2024llm}
\bibfield{author}{\bibinfo{person}{Nishat Raihan},
  \bibinfo{person}{Mohammed~Latif Siddiq}, \bibinfo{person}{Joanna Santos},
  {and} \bibinfo{person}{Marcos Zampieri}.} \bibinfo{year}{2024}\natexlab{}.
\newblock \showarticletitle{Large Language Models in Computer Science
  Education: A Systematic Literature Review}.
\newblock \bibinfo{journal}{\emph{arXiv preprint arXiv:2410.16349}}
  (\bibinfo{year}{2024}).
\newblock


\bibitem[Rasul et~al\mbox{.}(2012)]%
        {rasul2012employability}
\bibfield{author}{\bibinfo{person}{Mohamad~Sattar Rasul}, \bibinfo{person}{Rose
  Amnah~Abd Rauf}, \bibinfo{person}{Azlin~Norhaini Mansor}, {and}
  \bibinfo{person}{AP Puvanasvaran}.} \bibinfo{year}{2012}\natexlab{}.
\newblock \showarticletitle{Employability Skills Assessment Tool Development.}
\newblock \bibinfo{journal}{\emph{International Education Studies}}
  \bibinfo{volume}{5}, \bibinfo{number}{5} (\bibinfo{year}{2012}),
  \bibinfo{pages}{43--56}.
\newblock


\bibitem[Reichhart et~al\mbox{.}(2007)]%
        {reichhart2007rule}
\bibfield{author}{\bibinfo{person}{Stefan Reichhart}, \bibinfo{person}{Tudor
  G{\^\i}rba}, {and} \bibinfo{person}{St{\'e}phane Ducasse}.}
  \bibinfo{year}{2007}\natexlab{}.
\newblock \showarticletitle{Rule-based Assessment of Test Quality.}
\newblock \bibinfo{journal}{\emph{J. Object Technol.}} \bibinfo{volume}{6},
  \bibinfo{number}{9} (\bibinfo{year}{2007}), \bibinfo{pages}{231--251}.
\newblock


\bibitem[Ricchiardi and Emanuel(2018)]%
        {ricchiardi2018soft}
\bibfield{author}{\bibinfo{person}{Paola Ricchiardi} {and}
  \bibinfo{person}{Federica Emanuel}.} \bibinfo{year}{2018}\natexlab{}.
\newblock \showarticletitle{Soft skill assessment in higher education}.
\newblock \bibinfo{journal}{\emph{Journal of Educational, Cultural and
  Psychological Studies (ECPS Journal)}} \bibinfo{number}{18}
  (\bibinfo{year}{2018}), \bibinfo{pages}{21--53}.
\newblock


\bibitem[Shi and Huang(2019a)]%
        {shi2019el}
\bibfield{author}{\bibinfo{person}{Yaqing Shi} {and} \bibinfo{person}{Song
  Huang}.} \bibinfo{year}{2019}\natexlab{a}.
\newblock \showarticletitle{Research on Software Testing Technical Ability
  Training based on e-learning}. In \bibinfo{booktitle}{\emph{Proceedings of
  the 2019 4th International Conference on Distance Education and Learning}}.
  \bibinfo{pages}{30--34}.
\newblock


\bibitem[Shi and Huang(2019b)]%
        {shi2019research}
\bibfield{author}{\bibinfo{person}{Yaqing Shi} {and} \bibinfo{person}{Song
  Huang}.} \bibinfo{year}{2019}\natexlab{b}.
\newblock \showarticletitle{Research on Software Testing Technical Ability
  Training based on e-learning}. In \bibinfo{booktitle}{\emph{Proceedings of
  the 2019 4th International Conference on Distance Education and Learning}}.
  \bibinfo{pages}{30--34}.
\newblock


\bibitem[Song et~al\mbox{.}(2024)]%
        {song2024automated}
\bibfield{author}{\bibinfo{person}{Yishen Song}, \bibinfo{person}{Qianta Zhu},
  \bibinfo{person}{Huaibo Wang}, {and} \bibinfo{person}{Qinhua Zheng}.}
  \bibinfo{year}{2024}\natexlab{}.
\newblock \showarticletitle{Automated Essay Scoring and Revising Based on
  Open-Source Large Language Models}.
\newblock \bibinfo{journal}{\emph{IEEE Transactions on Learning Technologies}}
  (\bibinfo{year}{2024}).
\newblock


\bibitem[Sun et~al\mbox{.}(2019)]%
        {sun2019maf}
\bibfield{author}{\bibinfo{person}{Weisong Sun}, \bibinfo{person}{Xingya Wang},
  \bibinfo{person}{Haoran Wu}, \bibinfo{person}{Ding Duan},
  \bibinfo{person}{Zesong Sun}, {and} \bibinfo{person}{Zhenyu Chen}.}
  \bibinfo{year}{2019}\natexlab{}.
\newblock \showarticletitle{MAF: method-anchored test fragmentation for test
  code plagiarism detection}. In \bibinfo{booktitle}{\emph{2019 IEEE/ACM 41st
  International Conference on Software Engineering: Software Engineering
  Education and Training (ICSE-SEET)}}. IEEE, \bibinfo{pages}{110--120}.
\newblock


\bibitem[Tang et~al\mbox{.}(2024)]%
        {tang2024harnessing}
\bibfield{author}{\bibinfo{person}{Xiaoyi Tang}, \bibinfo{person}{Hongwei
  Chen}, \bibinfo{person}{Daoyu Lin}, {and} \bibinfo{person}{Kexin Li}.}
  \bibinfo{year}{2024}\natexlab{}.
\newblock \showarticletitle{Harnessing LLMs for multi-dimensional writing
  assessment: Reliability and alignment with human judgments}.
\newblock \bibinfo{journal}{\emph{Heliyon}} \bibinfo{volume}{10},
  \bibinfo{number}{14} (\bibinfo{year}{2024}).
\newblock


\bibitem[Tenbergen(2024)]%
        {tenbergen2024tool}
\bibfield{author}{\bibinfo{person}{Bastian Tenbergen}.}
  \bibinfo{year}{2024}\natexlab{}.
\newblock \showarticletitle{A Tool to Facilitate Calibrated Peer Reviews in
  Software Engineering Education}. In \bibinfo{booktitle}{\emph{2024 36th
  International Conference on Software Engineering Education and Training
  (CSEE\&T)}}. IEEE, \bibinfo{pages}{1--3}.
\newblock


\bibitem[Tian et~al\mbox{.}(2022)]%
        {tian2022software}
\bibfield{author}{\bibinfo{person}{Jingbai Tian}, \bibinfo{person}{Jianghao
  Yin}, {and} \bibinfo{person}{Liang Xiao}.} \bibinfo{year}{2022}\natexlab{}.
\newblock \showarticletitle{Software requirements engineer’s ability
  assessment method based on empirical software engineering}.
\newblock \bibinfo{journal}{\emph{Wireless Communications and Mobile
  Computing}} \bibinfo{volume}{2022}, \bibinfo{number}{1}
  (\bibinfo{year}{2022}), \bibinfo{pages}{3617140}.
\newblock


\bibitem[Towey and Chen(2015)]%
        {towey2015teaching}
\bibfield{author}{\bibinfo{person}{Dave Towey} {and}
  \bibinfo{person}{Tsong~Yueh Chen}.} \bibinfo{year}{2015}\natexlab{}.
\newblock \showarticletitle{Teaching software testing skills: Metamorphic
  testing as vehicle for creativity and effectiveness in software testing}. In
  \bibinfo{booktitle}{\emph{2015 IEEE International Conference on Teaching,
  Assessment, and Learning for Engineering (TALE)}}. IEEE,
  \bibinfo{pages}{161--162}.
\newblock


\bibitem[Ullah et~al\mbox{.}(2019)]%
        {ullah2019rule}
\bibfield{author}{\bibinfo{person}{Zahid Ullah}, \bibinfo{person}{Adidah
  Lajis}, \bibinfo{person}{Mona Jamjoom}, \bibinfo{person}{Abdulrahman~H
  Altalhi}, \bibinfo{person}{Jalal Shah}, {and} \bibinfo{person}{Farrukh
  Saleem}.} \bibinfo{year}{2019}\natexlab{}.
\newblock \showarticletitle{A rule-based method for cognitive competency
  assessment in computer programming using Bloom’s taxonomy}.
\newblock \bibinfo{journal}{\emph{IEEE Access}}  \bibinfo{volume}{7}
  (\bibinfo{year}{2019}), \bibinfo{pages}{64663--64675}.
\newblock


\bibitem[Wang et~al\mbox{.}(2019)]%
        {wang2019software}
\bibfield{author}{\bibinfo{person}{Xingya Wang}, \bibinfo{person}{Weisong Sun},
  \bibinfo{person}{Linghuan Hu}, \bibinfo{person}{Yuan Zhao},
  \bibinfo{person}{W~Eric Wong}, {and} \bibinfo{person}{Zhenyu Chen}.}
  \bibinfo{year}{2019}\natexlab{}.
\newblock \showarticletitle{Software-testing contests: Observations and lessons
  learned}.
\newblock \bibinfo{journal}{\emph{Computer}} \bibinfo{volume}{52},
  \bibinfo{number}{10} (\bibinfo{year}{2019}), \bibinfo{pages}{61--69}.
\newblock


\bibitem[Wieser et~al\mbox{.}(2023)]%
        {wieser2023investigating}
\bibfield{author}{\bibinfo{person}{Markus Wieser}, \bibinfo{person}{Klaus
  Sch{\"o}ffmann}, \bibinfo{person}{Daniela Stefanics},
  \bibinfo{person}{Andreas Bollin}, {and} \bibinfo{person}{Stefan Pasterk}.}
  \bibinfo{year}{2023}\natexlab{}.
\newblock \showarticletitle{Investigating the Role of ChatGPT in Supporting
  Text-Based Programming Education for Students and Teachers}. In
  \bibinfo{booktitle}{\emph{International Conference on Informatics in Schools:
  Situation, Evolution, and Perspectives}}. Springer Nature Switzerland Cham,
  \bibinfo{pages}{40--53}.
\newblock


\bibitem[Zhou et~al\mbox{.}(2022)]%
        {zhou2022meta}
\bibfield{author}{\bibinfo{person}{Tianqi Zhou}, \bibinfo{person}{Jiawei Liu},
  \bibinfo{person}{Yifan Wang}, {and} \bibinfo{person}{Zhenyu Chen}.}
  \bibinfo{year}{2022}\natexlab{}.
\newblock \showarticletitle{META: multidimensional evaluation of testing
  ability}. In \bibinfo{booktitle}{\emph{Proceedings of the ACM/IEEE 44th
  International Conference on Software Engineering: Companion Proceedings}}.
  \bibinfo{pages}{139--143}.
\newblock


\end{thebibliography}

\end{document}